\documentclass[12pt]{article}
\usepackage{amsmath,amssymb}
\usepackage{graphicx}
\newcommand{\eqdef}{\stackrel{\text{def}}{=}}

\newcommand{\n}{\nonumber \\}
\newcommand{\bm}{\boldsymbol}
\newcommand{\ignore}[1]{}

\numberwithin{equation}{section}
\newcommand{\Romannumeral}[1]{\uppercase\expandafter{\romannumeral#1}}

\newtheorem{theo}{\bf Theorem}[section]

\newtheorem{rema}[theo]{\bf Remark}

\newcommand{\ma}{\hspace{0pt}}

\allowdisplaybreaks[4]

\begin{document}

\baselineskip=20pt
\newcommand{\preprint}{
\vspace*{-20mm}\begin{flushleft}\end{flushleft}
}
\newcommand{\Title}[1]{{\baselineskip=26pt
  \begin{center} \Large \bf #1 \\ \ \\ \end{center}}}
\newcommand{\Author}{\begin{center}
  \large \bf 
  Ryu Sasaki${}$ \end{center}}
\newcommand{\Address}{\begin{center}
     Department of Physics and Astronomy, Tokyo University of Science,
     Noda 278-8510, Japan
        \end{center}}
\newcommand{\Accepted}[1]{\begin{center}
  {\large \sf #1}\\ \vspace{1mm}{\small \sf Accepted for Publication}
  \end{center}}

\preprint
\thispagestyle{empty}

\Title{Exactly solvable inhomogeneous fermion systems}

\Author

\Address
\vspace{1cm}

\begin{abstract}
15 exactly solvable inhomogeneous (spinless) fermion systems on one-dimensional lattices
are constructed explicitly based on the discrete orthogonal polynomials of Askey scheme,
{\em e.g.} the Krawtchouk, Hahn,  Racah,  Meixner,  $q$-Racah  polynomials.
The Schr\"odinger and Heisenberg equations are solved explicitly, as
the entire set of the eigenvalues and eigenstates are known explicitly.
The ground state two point correlation functions are derived explicitly.
The multi point correlation functions are obtained by Wick's  Theorem.
Corresponding 15 exactly solvable XX spin systems are also displayed.
They all have nearest neighbour interactions.
The exact solvability of Schr\"odinger equation means that of the corresponding Fokker-Planck equation.
This leads to 15 exactly solvable Birth and Death fermions and 15 Birth and Death spin models.
These provide plenty of materials for calculating interesting quantities, {\em e.g.} entanglement entropy, etc.
\end{abstract}

%
%
\section{Introduction}
\label{sec:intro}
Here I report a simple construction of exactly solvable inhomogeneous (spinless) fermions on a
one dimensional integer lattice. They all have nearest neighbour interactions.
Compared to the plethora of exactly solvable quantum mechanical systems 
\cite{susyqm, cal,sut, mos, gomez, gomez1, quesne, os16, os25},
the list of known solvable models of fermions on a lattice or continuum is remarkably short.
This is a modest attempt to add some to the list.
The principle of constructing exactly solvable inhomogeneous fermion systems on a 
one dimensional integer lattice is quite general and simple.
Pick up a hermitian or self-adjoint matrix $\mathcal{H}$, finite or semi-infinite dimensional,
for which {\em the complete set of discrete eigenvalues and corresponding orthonormal eigenvectors are known.}
The one dimensional integer lattice, finite or semi-infinite, is denote by $\mathcal{X}$,
and their points are displayed by $x$, $y$ and $z$ for analytic treatments,
\begin{align}
&x,y,z\in \mathcal{X}=\{0,1,\ldots,N\}, \quad N\in\mathbb{N},\quad x,y,z\in \mathcal{X}=\mathbb{Z}_{\ge0},
\label{lattice}\\
&\qquad \mathcal{H}=\bigl(\mathcal{H}_{x,y}\bigr),\quad 
\mathcal{H}_{x,y}^*=\mathcal{H}_{y,x}\in\mathbb{C},\qquad x,y\in\mathcal{X}.
\label{H0}
\end{align}
Without loss of generality, $\mathcal{H}$ is assumed to be positive semi-definite, by adjusting 
the diagonal components,
\begin{align}
\mathcal{H}\hat{\phi}_n&=\mathcal{E}(n)\hat{\phi}_n \Longleftrightarrow
\sum_{y\in\mathcal{X}}\mathcal{H}_{x,y}\hat{\phi}_n(y)=\mathcal{E}(n)\hat{\phi}_n(x),\quad
\mathcal{E}(n)\ge0,\ \hat{\phi}_n(x)\in\mathbb{C},\ x,n\in\mathcal{X},
\label{Heig0}\\
&\sum_{x\in\mathcal{X}}\hat{\phi}_m^*(x)\hat{\phi}_n(x)=\delta_{m,n},\quad 
\sum_{n\in\mathcal{X}}\hat{\phi}_n^*(x)\hat{\phi}_n(y)=\delta_{x,y},\quad x,y,m,n\in\mathcal{X}.
\end{align}
The inhomogenous spinless fermion Hamiltonian $\mathcal{H}_f$  on $\mathcal{X}$
is constructed fron $\mathcal{H}$ 
as a bi-linear form of the fermion operators $\{c_x\}$ and $\{c_x^\dagger\}$, 
 obeying the
canonical anti-commutation relations,
\begin{align}
 \{c_x^\dagger,c_y\}&=\delta_{x,y},\quad \{c_x^\dagger,c_y^\dagger\}=0=\{c_x,c_y\},
\quad x,y\in\mathcal{X},
\label{comrel}\\
\mathcal{H}_f&\eqdef \sum_{x,y\in\mathcal{X}}c_x^\dagger\mathcal{H}_{x,y}c_y.
\label{Hf0}
\end{align}
It is diagonalised by the momentum space fermion operators 
$\{\hat{c}_n\}$, $\{\hat{c}_n^\dagger\}$, $n\in\mathcal{X}$,
\begin{align}
\hat{c}_n&\eqdef\sum_{x\in\mathcal{X}}\hat{\phi}_n^*(x)c_x,\ \
\hat{c}_n^\dagger=\sum_{x\in\mathcal{X}}\hat{\phi}_n(x)c_x^\dagger\ \Leftrightarrow \
{c}_x=\sum_{n\in\mathcal{X}}\hat{\phi}_n(x)\hat{c}_n,\ \
{c}_x^\dagger=\sum_{m\in\mathcal{X}}\hat{\phi}_m^*(x)\hat{c}_m^\dagger,
\label{momrep}\\
&\hspace{2cm} \Longrightarrow \ \{\hat{c}_m^\dagger,\hat{c}_n\}
=\delta_{m\,n},\ \{\hat{c}_m^\dagger,\hat{c}_n^\dagger\}=0=\{\hat{c}_m,\hat{c}_n\},
\label{chatcom0}\\[-2pt]
&\hspace{6cm} \Downarrow \n[-2pt]
 &\mathcal{H}_f=\sum_{m,n,x,y\in\mathcal{X}}
\hat{\phi}_m^*(x)\mathcal{H}_{x,y}\hat{\phi}_n(y)\hat{c}_m^\dagger\hat{c}_n
=\sum_{m,n,x\in\mathcal{X}}\mathcal{E}(n)\hat{\phi}_m^*(x)\hat{\phi}_n(x)\hat{c}_m^\dagger\hat{c}_n\n
&\phantom{\mathcal{H}_f}=\sum_{n\in\mathcal{X}}\mathcal{E}(n)\hat{c}_n^\dagger\hat{c}_n,
\label{Hjdiag}\\
&\hspace{2cm} \Longrightarrow \ [\mathcal{H}_{f},\hat{c}_n^\dagger]=\mathcal{E}(n)\hat{c}_n^\dagger,
\qquad [\mathcal{H}_{f},\hat{c}_n]=-\mathcal{E}(n)\hat{c}_n.
\label{Hfcom10}
\end{align}

\bigskip
In this paper I present 15 real symmetric {\em tri-diagonal} Hamiltonians $\mathcal{H}$ with
the complete sets of eigensystems.
The eigenvalues are simple and the zero-mode of the Hamiltonian 
$\mathcal{H}\hat{\phi}_0=0$, $\mathcal{E}(0)=0$ is nodeless, 
$\hat{\phi}_0(x)>0$, $x\in\mathcal{X}$.
The 15 discrete orthogonal polynomials of Askey scheme \cite{askey,kls,ismail}
are the eigenvectors of these Hamiltonians $\mathcal{H}^A$ \cite{os12},
as explained in some detail in section two. The superscript $A$ is for Askey scheme.
These Hamiltonians produce lattice fermion Hamiltonians $\mathcal{H}_f$ \eqref{Hfdef1} 
with the nearest neighbour interactions
as explored in detail in section three.
As is well known these exactly solvable fermions are related to XX spin models 
by the Jordan-Wigner transformations. 
The basic structures of these 15 exactly solvable XX spin chains $\mathcal{H}_s$ \eqref{Hsdef1} are
surveyed in section four.
It is also well known that the Schr\"odinger equations are related by a similarity transformation 
to the Fokker-Planck equations \cite{risken} describing stochastic processes.
The stochastic processes with the nearest neighbour interactions are called Birth and Death (BD)
processes \cite{ismail,feller}.
The exactly solvable classical BD processes \cite{bdsol} related to $\mathcal{H}^A$ 
are briefly recapitulated in \S\ref{sec:BDeq}.
The quantum versions are the  15 exactly solvable quantum fermion walks 
\S\ref{sec:FBDeq} and quantum spin walks.
As shown in \eqref{strongdecay}, the multiple fermion walks 
and walks with multiple spin excitations decay rapidly.
Single fermion walks and single spin excitation walks show interesting features
 \cite{gvz,ortho,Kra,qcmarkov}.
 
 \bigskip
 This project is deeply influenced and catalysed by the seminal works of Gr\"unbaum, Nepomecie, Vinet, Zhedanov 
 and their coauthors \cite{gvz,ortho,Kra}. 
 They introduced exactly solvable nearest neighbour interaction  fermions, quantum fermion walks 
 and quantum spin walks through the classical BD processes \`a la Karlin-McGregor \cite{KarMcG,KarMcG2}.
 That is, the coordinates of fermions are $m,n,\ldots$ and
 they used the three term recurrence relations of the discrete orthogonal polynomials 
 \begin{equation*}
 A_n\bigl(P_{n+1}(x)-P_n(x)\bigr)+C_n\bigl(P_{n-1}(x)-P_n(x)\bigr)= x P_n(x),
\end{equation*}
 for the description of the nearest neighbour interactions. 
 Here $x$ corresponds to the momentum (dual) label and the polynomial $P_n(x)$ is of degree $n$ in $x$.
 
  In contrast, in my case $x$ is for the coordinate of the fermions and nearest neighbour interactions are
 taken care of by the difference equation \eqref{difeq} \cite{os12,bdsol}. The labels $m,n$ are for the momentum and
 the energy $\mathcal{E}(n)$ of the exactly solvable theories takes five different forms as shown in \eqref{En}.
 My theory covers full 15 polynomials as demonstrated here.
 
 In Karlin-McGregor prescription, the polynomial $P_n(x)$ in the momentum (dual) space
 has a linear eigenvalue, {\em i.e.} type (i) in \eqref{En}.
 That is why, their method is usable, within exactly solvable theories,  
 only for four polynomials, the Krawtchouk (K), dual Hahn (dH), Meixner (M) and Charlier (C).

Many explicit examples of exactly solvable theories obtained here 
allow to evaluate various interesting physical
quantities, {\em e.g.} entanglement entropy, etc \cite{gvz,ortho,Kra,finkel,latorre}.
It would be amusing to see how the entropies depend on $0<q<1$
 in  theories involving  $q$-polynomials.
 
\section{Discrete orthogonal polynomials of Askey scheme}
\label{sec:disc}

In order to prepare the necessary notions and notation used in the subsequent sections,
let us start with a brief review of the discrete orthogonal polynomials of Askey scheme \cite{askey,kls},
excluding those having the Jackson integral measures \cite{askey}.
The fifteen polynomials are,  the Krawtchouk (K), Hahn (H), dual Han (dH), Racah (R), Meixner (M), Chrlier (C),
quantum $q$-Krawtchouk (q$q$K), $q$-Krawtchouk ($q$K), affine $q$-Krawtchouk (a$q$K),
$q$-Hahn ($q$H), dual $q$-Hahn (d$q$H) and $q$-Racah ($q$R),  little $q$-Jacobi (l$q$J), 
little $q$-Laguerre (l$q$L), Al-Salam-Carlitz II (ASC) polynomials.
They are defined on a one dimensional integer lattice $\mathcal{X}$, finite or semi-infinite \eqref{lattice},
and their points are displayed by $x$, $y$ and $z$ for analytic treatments.
A discrete orthogonal polynomial $\check{P}_n(x)=P_n\bigl(\eta(x)\bigr)$, $x,n\in\mathcal{X}$ 
is a degree $n$ polynomial in $\eta(x)$. 
The {\em sinusoidal coordinate} $\eta(x)$ is a linear or quadratic function of $x$ or $q^{\pm x}$ ($0<q<1$) which
vanishes at $x=0$, $\eta(0)=0$ \cite{os12} and takes one of the following five forms,
\begin{equation}
   \begin{split}
  &\text{(\romannumeral1)}\ \eta(x)=x,\  K, H, M, C,\quad  \text{(\romannumeral2)}\ \eta(x)=x(x+d),\  dH, R \\
&\text{(\romannumeral3)}\ \eta(x)=1-q^x,\ lqJ, lqL\qquad \text{(\romannumeral4)}
\ \eta(x)=q^{-x}-1,\ qH, qqK, qK, aqK, ASC, \\
&\text{(\romannumeral5)}\ \eta(x)=(q^{-x}-1)(1-d q^x),\quad dqH, qR.
  \end{split}
  \label{5eta}
\end{equation}
They are all increasing functions of $x$.
\subsection{Difference equation $\widetilde{\mathcal H}$}
\label{sec:difeq}
These discrete orthogonal polynomials all satisfy, on top of the well-known three term recurrence relations,
 the following type of difference equations \cite{kls,os12},
\begin{align}
&B(x)\bigl(\check{P}_n(x)-\check{P}_n(x+1)\bigr)+D(x)\bigl(\check{P}_n(x)-\check{P}_n(x-1)\bigr)
=\mathcal{E}(n)\check{P}_n(x),\quad x,n\in\mathcal{X},
\label{difeq}\\
&\hspace{7cm} \Downarrow \n
&\hspace{4.4cm}\sum_{y\in\mathcal{X}} \widetilde{\mathcal{H}}_{x,y}\check{P}_n(y)
=\mathcal{E}(n)\check{P}_n(x),
\label{Hteq}\\
 &\widetilde{\mathcal{H}}=(\widetilde{\mathcal{H}}_{x,y}),\quad
  \widetilde{\mathcal{H}}_{x,y}\eqdef B(x)(\delta_{x,y}-\delta_{x+1,y})+
  D(x)(\delta_{x,y}-\delta_{x-1,y}),
  \label{Htdef}
\end{align}
in which $\mathcal{E}(n)\ge0$ is the eigenvalue and the positive coefficient functions $B(x)$ and $D(x)$
satisfy the boundary conditions,
\begin{equation}
B(x)>0,\ D(x)>0,\ x\in\mathcal{X},\quad D(0)=0,\quad B(N)=0, \quad (\text{finite case}).
\label{bound}
\end{equation}
The eigenvalue $\mathcal{E}(n)$ is a linear or quadratic function of $n$ or $q^{\pm n}$ which
vanishes at $n=0$, $\mathcal{E}(0)=0$ \cite{os12} and increases with $n$,
\begin{equation}
   \begin{split}
  &\text{(\romannumeral1)}\ \mathcal{E}(n)=n,\  K, dH, M, C,
  \quad  \text{(\romannumeral2)}\ \mathcal{E}(n)=n(n+d),\  H, R, \\
&\text{(\romannumeral3)}\ \mathcal{E}(n)=1-q^n,\ qqK, ASC,\qquad \text{(\romannumeral4)}
\ \mathcal{E}(n)=q^{-n}-1,\ dqH, aqK, lqL\\
&\text{(\romannumeral5)}\ \mathcal{E}(n)=(q^{-n}-1)(1-d q^n),\quad qK, qH, qR, lqJ.
  \end{split}
  \label{En}
\end{equation}  
It should be stressed that the coefficients $B(x)$ and $D(x)$ are independent of the changes 
of the normalisation of the  polynomials of various degrees, whereas the changes 
 force the redefinition of the coefficients of the
three term recurrence relations.

\subsection{Hamiltonian $\mathcal{H}^A$}
\label{sec:Ham}

By introducing a positive function $\phi_0(x)$, $x\in\mathcal{X}$, defined by the ratios of $B(x)$ and $D(x+1)$,
\begin{align} 
&\frac{\phi_0(x+1)}{\phi_0(x)}\eqdef\sqrt{\frac{B(x)}{D(x+1)}},\quad \phi_0(x)>0,\quad
x\in\mathcal{X}, \quad \phi_0(0)\eqdef1,
\label{phi0def}\\
&\qquad \Longrightarrow \phi_0(x)=\prod_{y=0}^{x-1}\sqrt{\frac{B(y)}{D(y+1)}},
\label{phi0def2}
\end{align}
the above difference operator, $\widetilde{\mathcal H}^A$ \eqref{Htdef} is transformed to 
a real symmetric tri-diagonal matrix (Hamiltonian) $\mathcal{H}^A$,
\begin{align} 
\mathcal{H}^A_{x,y}&\eqdef \phi_0(x)\widetilde{\mathcal H}^A_{x,y}\phi_0(y)^{-1}
=\mathcal{H}^A_{y,x},\qquad \qquad \qquad x,y\in\mathcal{X},
\label{Hdef1}\\
&=\phi_0(x)\bigl(B(x)(\delta_{x,y}-\delta_{x+1,y})+
  D(x)(\delta_{x,y}-\delta_{x-1,y})\bigr)\phi(y)^{-1}\n
&=\bigl(B(x)+D(x)\bigr)\delta_{x,y}-\phi_0(x)\frac{B(x)}{\phi_0(x+1)}\delta_{x+1,y}-\phi_0(x)\frac{D(x)}{\phi_0(x-1)}\delta_{x-1,y}\n
&=\bigl(B(x)+D(x)\bigr)\delta_{x,y}-\sqrt{B(x)D(x+1)}\delta_{x+1,y}-\sqrt{B(x-1)D(x)}\delta_{x-1,y},
\label{Hdef2}
\end{align}
{\scriptsize
\begin{equation*}
\mathcal{H}^A=\left(
\begin{array}{cccccc}
\!\!B(0)  & -\sqrt{B(0)D(1)}  &   0& \cdots&\cdots&0\\
\!\!-\sqrt{B(0)D(1)}  & B(1)+D(1)  & -\sqrt{B(1)D(2)} &0&\cdots&\vdots  \\
\!\!0  &  -\sqrt{B(1)D(2)}  &   B(2)+D(2)&-\sqrt{B(2)D(3)}&\cdots&\vdots\\
\!\!\vdots&\cdots&\cdots&\cdots&\cdots&\vdots\\
\!\!\vdots&\cdots&\cdots&\cdots&\cdots&0\\
\!\!0&\cdots&\cdots&-\sqrt{\!B(N\!\!-\!\!2)D(N\!\!-\!\!1)}&B(N\!\!-\!\!1)\!
+\!D(N\!\!-\!\!1)&-\sqrt{\!B(N\!\!-\!\!1)D(N)}\\
\!\!0&\cdots&\cdots&0&-\sqrt{B(N\!-\!1)D(N)}&D(N)
\end{array}
\right).
\end{equation*}
}
The difference equation for the polynomials $\{\check{P}_n(x)=P_n\bigl(\eta(x)\bigr)\}$ \eqref{difeq}, \eqref{Hteq} is now 
an eigenvalue problem of the tri-diagonal Hamiltonian $\mathcal{H}^A$ \eqref{Hdef1}
\begin{equation}
\mathcal{H}^A\phi_n(x)=\mathcal{E}(n)\phi_n(x),\quad \sum_{y\in\mathcal{X}}\mathcal{H}^A_{x,y}\phi_n(y)=\mathcal{E}(n)\phi_n(x),
\quad \phi_n(x)\eqdef\phi_0(x)\check{P}_n(x),\quad 
x,n\in\mathcal{X},
\label{Heq}
\end{equation}
which describes {\em the nearest neighbour interactions.}
Since  the eigenvalues of a tri-diagonal Hamiltonians are simple, the eigenvectors $\{\phi_n(x)\}$, $n\in\mathcal{X}$
are mutually orthogonal.
It is known that the Hamiltonia $\mathcal{H}^A$ \eqref{Hdef1} is {\em positive semi-definite} $\mathcal{E}(n)\ge0$,
see \cite{os12}(2.12).
By introducing the universal normalisation of the discrete polynomials,
\begin{equation}
\check{P}_n(0)=P_n\bigl(\eta(0)\bigr)=P_n(0)=1,\quad n\in\mathcal{X},\qquad \check{P}_0(x)\equiv1,
\label{univn}
\end{equation}
the orthogonality relations of the polynomials $\{\check{P}_n(x)\}$ read
\begin{equation}
\sum_{x\in\mathcal{X}}\phi_m(x)\phi_n(x)=\sum_{x\in\mathcal{X}}\phi_0^2(x)\check{P}_m(x)\check{P}_n(x)=
\frac{\delta_{m,n}}{d_n^2},\quad d_n>0,\quad m,n\in\mathcal{X}.
\label{orthrel}
\end{equation}
As the first component of the eigenvectors of a tri-diagonal Hamiltonian is non-vanishing, the above
normalisation \eqref{univn} is always possible.
The positive function $\phi_0(x)$ is the zero-mode of the Hamiltonian $\mathcal{H}^A$ \eqref{Hdef2},
\begin{equation*}
\mathcal{H}^A\phi_0(x)=0,\quad \mathcal{E}(0)=0,
\end{equation*}
and  it is the square root of the orthogonality measure $\phi_0^2(x)$ of the polynomials $\{\check{P}_n(x)\}$.
The set of orthonormal eigenvectors $\{\hat{\phi}_n(x)\}$, $n\in\mathcal{X}$ is 
\begin{align}    
&\hat{\phi}_n(x)\eqdef d_n\phi_n(x)=d_n\phi_0(x)\check{P}_n(x)=d_n\phi_0(x)P_n\bigl(\eta(x)\bigr),
\quad n,x\in\mathcal{X},
\label{hatdef}\\
&\sum_{x\in\mathcal{X}}\hat{\phi}_m(x)\hat{\phi}_n(x)=\delta_{m,n},\quad
\sum_{n\in\mathcal{X}}\hat{\phi}_n(x)\hat{\phi}_n(y)=\delta_{x,y},
\label{ortrel2}
\end{align}
in which the latter means the completeness of the eigenfunctions.

Here is a summary of the exact solvability of the 15 discrete orthogonal polynomials of Askey scheme.
The polynomial $\check{P}_n(x)=P_n\bigl(\eta(x)\bigr)$ is a terminating hypergeometric function and is a degree
$n$ polynomial in the sinusoidal coordinate $\eta(x)$ \eqref{5eta}, 
satisfying the difference equation \eqref{difeq} with the eigenvalue $\mathcal{E}(n)$ \eqref{En}.  
The function $\phi_0(x)$ \eqref{phi0def}, the square root of the orthogonality weight,
and the normalisation constant \eqref{orthrel} have explicit closed form expressions.
They are all explicitly known and they  play an essential role to diagonalise the fermion Hamiltonian 
$H_f$ \eqref{Hfdef1} introduced
shortly.

\subsection{Classical Birth and Death operator $\mathcal{L}_{BD}$}
\label{sec:BDeq}

It is well known that Schr\"odinger equations are closely related to Fokker-Planck equations \cite{risken}.
If one is exactly solvable, so is the other.
The Fokker-Planck equation corresponding to the Hamiltonian $\mathcal{H}^A$ \eqref{Hdef1} of the
discrete orthogonal polynomial is called Birth and Death (BD) equation \cite{ismail}\S5.2,
 due to the nearest neighbour interactions.
 They are exactly solvable \cite{bdsol}.
 Corresponding to \eqref{Hdef1}, the Birth and Death operator $\mathcal{L}_{BD}$ is defined by
 \begin{align}
(\mathcal{L}_{BD})_{x,y}&\eqdef -\phi_0(x)\mathcal{H}^A_{x,y}\phi_0(y)^{-1},\quad x,y\in\mathcal{X},
\label{LBDdef1}\\
&=B(x-1)\delta_{x-1,y}-B(x)\delta_{x,y}+
  D(x+1)\delta_{x+1,y}-D(x)\delta_{x,y}, 
  \label{LBDdef2}\\
\sum_{x\in\mathcal{X}}(\mathcal{L}_{BD})_{x,y}&=0.
\label{zerosum}
\end{align}
The BD equation for the probability distribution $\mathcal{P}(x;t)\ge0$ reads
\begin{equation}
\frac{\partial \mathcal{P}(x;t)}{\partial t}=\sum_{y\in\mathcal{X}}(\mathcal{L}_{BD})_{x,y}\mathcal{P}(y;t),
\quad \sum_{x\in\mathcal{X}}\mathcal{P}(x;t)=1,\quad x\in \mathcal{X}.
\label{BDeq}
\end{equation}
The conservation of the probability $\sum_{x\in\mathcal{X}}\mathcal{P}(x;t)=1$ is the consequence of the
zero sum property of $\mathcal{L}_{BD}$ \eqref{zerosum} and the initial condition 
$\sum_{x\in\mathcal{X}}\mathcal{P}(x;0)=1$.
By construction, $\{\hat{\phi}_0(x)\hat{\phi}_n(x)\}$, $n\in\mathcal{X}$ is the complete set of eigenvectors
of $\mathcal{L}_{BD}$,
\begin{equation}
\sum_{y\in\mathcal{X}}(\mathcal{L}_{BD})_{x,y}\hat{\phi}_0(y)\hat{\phi}_n(y)
=-\mathcal{E}(n)\hat{\phi}_0(x)\hat{\phi}_n(x),
\quad x\in\mathcal{X},
\label{LBDeig}
\end{equation}
which is {\em negative semi-definite}.
The solution of the BD equation \eqref{BDeq}  with the initial condition $\mathcal{P}(x;0)=\delta_{x,y}$ is
\cite{bdsol}
\begin{align}
\mathcal{P}(x,y;t)&=\hat{\phi}_0(x)\hat{\phi}_0(y)^{-1}
\sum_{n\in\mathcal{X}}e^{-\mathcal{E}(n)t}\hat{\phi}_n(x)\hat{\phi}_n(y),
\quad
t>0,
\label{bdsoly1}\\
\lim_{t\to+\infty}\mathcal{P}(x,y;t)&=\hat{\phi}_0(x)^2,
\label{limt}
\end{align}
and the solution corresponding to an arbitrary initial condition can be constructed by their linear combination.
 
\section{Exactly solvable inhomogeneous free fermions associated with the discrete orthogonal  polynomials}
\label{sec:solvfermi}
Here I present fifteen exactly solvable inhomogeneous free fermion systems on a 
one dimensional integer lattice \eqref{lattice} with the lattice points being denoted by $x,y,z\in\mathcal{X}$,
as in one dimensional field theory.
The main idea is to construct a fermion version  of the exactly solvable nearest neighbour interaction 
Hamiltonian $\mathcal{H}^A$ \eqref{Hdef2}.
\subsection{Exactly solvable Hamiltonian}
\label{sec:solvHam}
Choose one of the 15 discrete orthogonal polynomials $\{\check{P}_n(x)\}$ and construct 
a fermion Hamiltonian $H_f$ \eqref{Hfdef1} by picking up the
corresponding coefficient functions $B(x)$ and $D(x)$ of the difference equation \eqref{Htdef}.
The conventionally chosen form is
\begin{equation}
\mathcal{H}_{f}=\sum_{x\in\mathcal{X}}\left\{-\sqrt{B(x)D(x+1)}\bigl(c_x^\dagger c_{x+1}+c_{x+1}^\dagger c_x\bigr)
+\bigl(B(x)+D(x)-\mu\bigr)c_x^\dagger c_x\right\},
\label{Hfdef1}
\end{equation}
in which $\mu$ is the chemical potential and $\{c_x\}$ and $\{c_x^\dagger\}$, $x\in\mathcal{X}$  
are one-dimensional spinless fermion operators obeying the
canonical anti-commutation relations \eqref{comrel}.
It is trivial to verify that the total fermion number is conserved by the Hamiltonian $\mathcal{H}_{f}$
\begin{equation}
[\mathcal{H}_{f}, \mathcal{F}]=0,\quad [\mathcal{F}, c_x]=-c_x,\quad [\mathcal{F}, c_x^\dagger]=c_x^\dagger,
\quad \mathcal{F}\eqdef\sum_{x\in\mathcal{X}}c_x^\dagger c_x.
\label{fcons}
\end{equation}
The vacuum $|0\rangle$ is defined by 
\begin{equation}
c_x|0\rangle=0,\quad \forall x\in\mathcal{X},\quad \langle0|0\rangle=1,\quad \mathcal{H}_f|0\rangle=0,
\label{vacdef}
\end{equation}
and the state space (Hilbert space) of $\mathcal{H}_f$ \eqref{Hfdef1}  consists of vectors
\begin{equation}
\prod_{x\in\mathcal{J}}c_x^\dagger|0\rangle,
\label{genstates}
\end{equation}
in which $\mathcal{J}$ is a subset of $\mathcal{X}$ consisting of mutually distinct elements.

\bigskip
Note that, in the finite lattice, the unwanted operators $c_{N+1}$ and $c_{N+1}^\dagger$ do not appear 
in $\mathcal{H}_f$ \eqref{Hfdef1} due to the boundary condition $B(N)=0$ \eqref{bound}.
In this connection, the second term of the fermion Hamiltonian $\mathcal{H}_f$ \eqref{Hfdef1} 
\begin{equation*}
-\sum_{x\in\mathcal{X}}\sqrt{B(x)D(x+1)}\,c_{x+1}^\dagger c_x
\end{equation*}
can be rewritten by $y\eqdef x+1$ as
\begin{equation*}
-\sum_{y\in\mathcal{X}\backslash\{0\}}\sqrt{B(y-1)D(y)}\,c_y^\dagger c_{y-1}.
\end{equation*}
The $y=0$ term can be added as $D(0)=0$ \eqref{bound} 
and the fermion Hamiltonian $\mathcal{H}_f$ \eqref{Hfdef1}
is now  closely related to the Hamiltonian $\mathcal{H}^A$ \eqref{Hdef2} of the discrete orthogonal polynomial,
\begin{align} 
\mathcal{H}_f&=\sum_{x\in\mathcal{X}}\Bigl\{-\sqrt{B(x)D(x+1)}\,c_x^\dagger c_{x+1}
-\sqrt{B(x-1)D(x)}\,c_{x}^\dagger c_{x-1}\n
&\qquad  \quad 
\quad +\bigl(B(x)+D(x)-\mu\bigr)c_x^\dagger c_x\Bigr\}
\label{Hfdef2}\\
&=\sum_{x,y\in\mathcal{X}}\mathcal{H}^A_{x,y}c_x^\dagger c_y-\mu\sum_{x\in\mathcal{X}}c_x^\dagger c_x.
\label{Hfdef3}
\end{align}
Thus  the general form of fermion Hamiltonian $\mathcal{H}_f$ \eqref{Hf0} is recovered with the extra
chemical potential term.
This one has the nearest neighbour interactions  corresponding to
the discrete orthogonal polynomials of Askey scheme.

The Schr\"odinger equation reads
\begin{equation}
i\frac{\partial|\psi(t)\rangle}{\partial t}=\mathcal{H}_f|\psi(t)\rangle,
\label{Scheq}
\end{equation}
in which $|\psi(t)\rangle$ is a state vector.

By changing the sign of the fermion operators according to the parity of the lattice points in $\mathcal{X}$,
\begin{equation}
c_x'\eqdef (-1)^xc_x,\quad {c_x'}^\dagger =(-1)^xc_x^\dagger, \quad x\in\mathcal{X},
\label{parch}
\end{equation}
the alternative fermion Hamiltonian $\mathcal{H}_{f'}$ \eqref{Hfpdef}  is obtained
\begin{equation}
\mathcal{H}_{f'}=\sum_{x\in\mathcal{X}}
\left\{\sqrt{B(x)D(x+1)}\bigl({c_x'}^\dagger c_{x+1}'+{c_{x+1}'}^\dagger c_x'\bigr)
+\bigl(B(x)+D(x)-\mu\bigr){c_x'}^\dagger c_x'\right\},
\label{Hfpdef}
\end{equation}
see \cite{ortho}(7.6), \cite{Kra}(2.2), \cite{finkel}(2.3)(7.2).
It  shares the solvability with the original Hamiltonian $\mathcal{H}_f$ \eqref{Hfdef1}.

Let us introduce the momentum lattice fermion operators 
$\{\hat{c}_n\}$, $\{\hat{c}_n^\dagger\}$, $n\in\mathcal{X}$ \eqref{momrep}
in terms of the orthonormal eigenvectors $\{\hat{\phi}_n(x)\}$ \eqref{hatdef} 
of the corresponding Hamiltonian $\mathcal{H}^A$ \eqref{Hdef2} of the discrete orthogonal polynomials
 ({\em the generalised Fourier transforms}),
\begin{align}
&\hat{c}_n\eqdef \sum_{x\in\mathcal{X}}\hat{\phi}_n(x)c_x, \quad
\hat{c}_n^\dagger= \sum_{x\in\mathcal{X}}\hat{\phi}_n(x)c_x^\dagger, \quad
[\mathcal{F},\hat{c}_n]=-\hat{c}_n,\quad [\mathcal{F},\hat{c}_n^\dagger]=\hat{c}_n^\dagger,
\quad  \ n\in\mathcal{X},
\label{chatdef}\\
&\{\hat{c}_m^\dagger,\hat{c}_n\}=\delta_{m\,n},\ 
\{\hat{c}_m^\dagger,\hat{c}_n^\dagger\}=0=\{\hat{c}_m,\hat{c}_n\},
\label{chatcom}\\
&\hspace{3cm}\Rightarrow c_x=\sum_{n\in\mathcal{X}}\hat{\phi}_n(x)\hat{c}_n,\quad 
c_x^\dagger=\sum_{n\in\mathcal{X}}\hat{\phi}_n(x)\hat{c}_n^\dagger,\quad
\quad x\in\mathcal{X},
\label{cxexp}\\
&\qquad \{\hat{c}_n,c_x^\dagger\}=\hat{\phi}_n(x)=\{\hat{c}_n^\dagger,c_x\},
\quad \{\hat{c}_n,c_x\}=0=\{\hat{c}_n^\dagger,c_x^\dagger\}.
\label{cchat}
\end{align}
This leads to the following
\begin{theo}
\label{thep:main}
The diagonalisation of the fermion Hamiltonian $\mathcal{H}_f$ \eqref{Hfdef1}, \eqref{Hfdef3},
\begin{align}
&\hspace{4cm} \mathcal{H}_{f}
=\sum_{n\in\mathcal{X}}\bigl(\mathcal{E}(n)-\mu\bigr)\hat{c}_n^\dagger \hat{c}_n,
\label{Hfdef4}\\
&
\quad [\mathcal{H}_{f},\hat{c}_n^\dagger]=\bigl(\mathcal{E}(n)-\mu\bigr)\hat{c}_n^\dagger,
\qquad [\mathcal{H}_{f},\hat{c}_n]=-\bigl(\mathcal{E}(n)-\mu\bigr)\hat{c}_n,
\label{Hfcom1}\\[2pt]
&
\quad [\mathcal{H}_{f},\prod_{j\in\mathcal{J}}\hat{c}_j^\dagger]
=\sum_{j\in\mathcal{J}}\bigl(\mathcal{E}(j)-\mu\bigr)\cdot\prod_{j\in\mathcal{J}}\hat{c}_j^\dagger,
\qquad [\mathcal{H}_{f},\prod_{j\in\mathcal{J}}\hat{c}_j]
=-\sum_{j\in\mathcal{J}}\bigl(\mathcal{E}(j)-\mu\bigr)\cdot\prod_{j\in\mathcal{J}}\hat{c}_j
\label{HfcomJ}\\[2pt]
&\quad  [\mathcal{H}_{f},c_x^\dagger]
=\sum_{m\in\mathcal{X}}\bigl(\mathcal{E}(m)-\mu\bigr)\hat{\phi}_m(x)\hat{c}_m^\dagger,
\quad  [\mathcal{H}_{f},c_x]
=-\sum_{m\in\mathcal{X}}\bigl(\mathcal{E}(m)-\mu\bigr)\hat{\phi}_m(x)\hat{c}_m,
\label{Hfcom2}
\end{align}
is achieved by plugging in the expressions of $\{c_x\}$ and $\{c_x^\dagger\}$ \eqref{cxexp} in terms 
of the momentum  lattice fermion operators $\{\hat{c}_n\}$ and $\{\hat{c}_n^\dagger\}$ \eqref{chatdef}
and using the eigenvalue  equation of $\hat{\phi}_n(x)$ \eqref{Heq}.
Here, $\mathcal{J}$ is a subset of $\mathcal{X}$ consisting of mutually distinct elements.
The commutation relations \eqref{Hfcom1} show that the Heisenberg equations of motion for the 
operators $\hat{c}_n^\dagger$ and $\hat{c}_n$ are solved 
\begin{equation}
\hat{c}_n^\dagger(t)=e^{i(\mathcal{E}(n)-\mu)t}\,\hat{c}_n^\dagger(0),\quad
\hat{c}_n(t)=e^{-i(\mathcal{E}(n)-\mu)t}\,\hat{c}_n(0),
\label{heisol}
\end{equation}
and the orthonormalisation of the operators $\hat{c}_n^\dagger$ and $\hat{c}_n$ in the Krylov subspace 
{\rm \cite{parker,kriveri}} stops at the first level.
\end{theo}
In fact, by inserting $c_x$ and $c_x^\dagger$ \eqref{cxexp} 
into the fermion Hamiltonian $\mathcal{H}_f$ \eqref{Hfdef2},
it  reads
\begin{align*} 
\mathcal{H}_f&=\sum_{m,n,x\in\mathcal{X}}\hat{\phi}_m(x)\left\{
-\sqrt{B(x)D(x+1)}\,\hat{\phi}_n(x+1)-\sqrt{B(x-1)D(x)}\,\hat{\phi}_n(x-1)\right.\\
&\hspace{3.5cm}\left. +\bigl(B(x)+D(x)-\mu\bigr)\hat{\phi}_n(x)\right\}
\hat{c}_m^\dagger\hat{c}_n\\
&=\sum_{m,n,x\in\mathcal{X}}\bigl(\mathcal{E}(n)-\mu\bigr)
\hat{\phi}_m(x)\hat{\phi}_n(x)\hat{c}_m^\dagger\hat{c}_n\\
&=\sum_{n\in\mathcal{X}}\bigl(\mathcal{E}(n)-\mu\bigr)\hat{c}_n^\dagger\hat{c}_n,
\end{align*}
in which the eigenvalue equation for $\hat{\phi}_n(x)$ \eqref{Heq} and the orthogonality relation \eqref{ortrel2}
are used.

The general eigenstate (eigenvector) of $\mathcal{H}_f$ \eqref{Hfdef1}, with the fermion number $|\mathcal{J}|$  has the form 
\begin{align}
\prod_{j\in\mathcal{J}}\hat{c}_j^\dagger|0\rangle,\quad 
\mathcal{H}_f\prod_{j\in\mathcal{J}}\hat{c}_j^\dagger|0\rangle
&=E\prod_{j\in\mathcal{J}}\hat{c}_j^\dagger|0\rangle,
\quad E=\sum_{j\in\mathcal{J}}\bigl(\mathcal{E}(j)-\mu\bigr),
\label{eigst}\\
 \mathcal{F}\prod_{j\in\mathcal{J}}\hat{c}_j^\dagger|0\rangle
 &=|\mathcal{J}|\prod_{j\in\mathcal{J}}\hat{c}_j^\dagger|0\rangle,
\label{fnumJ}
\end{align}
in which $\mathcal{J}$ is a subset of $\mathcal{X}$ consisting of mutually distinct elements.
The first excited state $\hat{c}_0^\dagger|0\rangle=d_0\sum_{x\in\mathcal{X}}\phi_0(x)c_x^\dagger|0\rangle$
has all the fermions at $x\in\mathcal{X}$ have the same sign.

\begin{rema}
\label{primefermi}
The situation is rather different for the other fermion Hamiltonian $\mathcal{H}_{f'}$ \eqref{Hfpdef}.
It is diagonalised as
\begin{equation}
\mathcal{H}_{f'}=\sum_{n\in\mathcal{X}}\bigl(\mathcal{E}(n)-\mu\bigr)\hat{c}_n^\dagger \hat{c}_n,
\quad \hat{c}_n\eqdef \sum_{x\in\mathcal{X}}(-1)^x\hat{\phi}_n(x)c'_x, \qquad
\hat{c}_n^\dagger= \sum_{x\in\mathcal{X}}(-1)^x\hat{\phi}_n(x){c'}_x^\dagger,
\quad 
\end{equation}
and the first excited state 
$\hat{c}_0^\dagger|0\rangle=d_0\sum_{x\in\mathcal{X}}(-1)^x\phi_0(x){c'}_x^\dagger|0\rangle$
has parity oscillating signs and other higher order excitations, too.
These facts are related to the choice of the homogeneous free fermion chain with a negative sign
\begin{equation*}
\mathcal{H}_{ff}=-\frac12\sum_{x=0}^{N-1}\bigl(c_{x+1}^\dagger c_x+c_x^\dagger c_{x+1}\bigr).
\end{equation*}
\end{rema}

The general single particle excitation solution of the Schr\"odinger equation \eqref{Scheq} reads
\begin{equation}
|\psi(t)\rangle=\sum_{n\in\mathcal{X}}\beta_n e^{-i(\mathcal{E}(n)-\mu)t}\hat{c}_n^\dagger|0\rangle
=\sum_{x,n\in\mathcal{X}}\beta_n\hat{\phi}_n(x)e^{-i(\mathcal{E}(n)-\mu)t}c_x^\dagger|0\rangle,
\quad \beta_n\in\mathbb{C}.
\label{singex}
\end{equation}
The special choice,  $\beta_n=\hat{\phi}_n(y)$, $\forall n\in\mathcal{X}$,
gives a solution  corresponding to the initial condition $|\psi_y(0)\rangle=c_y^\dagger|0\rangle$,
\begin{equation}
|\psi_y(t)\rangle=\sum_{x,n\in\mathcal{X}}\hat{\phi}_n(y)
\hat{\phi}_n(x)e^{-i(\mathcal{E}(n)-\mu)t}c_x^\dagger|0\rangle.
\label{singex1}
\end{equation}
The transition amplitude between one fermion state 
$c_y^\dagger|0\rangle$ at $t=0$ and $c_x^\dagger|0\rangle$ at time $t$
is
\begin{equation}
\langle0|c_x|\psi_y(t)\rangle=\sum_{n\in\mathcal{X}}\hat{\phi}_n(y)\hat{\phi}_n(x)e^{-i(\mathcal{E}(n)-\mu)t}.
\label{1yx}
\end{equation}

The general $|\mathcal{J}|$ particle excitation solution of the Schr\"odinger equation \eqref{Scheq} is a linear 
combination of 
\begin{equation}
|\psi_\mathcal{J}(t)\rangle= e^{-i\sum_{j\in\mathcal{J}}(\mathcal{E}(n)-\mu)t}
\prod_{j\in\mathcal{J}}\hat{c}_j^\dagger|0\rangle,
\label{fJgex}
\end{equation}
in which $\mathcal{J}$ is a subset of $\mathcal{X}$ consisting of distinct elements.
%
%
\subsection{Ground state and correlation functions }
\label{sec:ground}
In this subsection, let us assume that the chemical potential is positive $\mu>0$ and
$\mu<\mathcal{E}(N)$ in the finite case. An integer $K\in\mathcal{X}$ is fixed by the condition,
\begin{equation}
\mathcal{E}(K)<\mu<\mathcal{E}(K+1),\qquad \mathcal{K}\eqdef\{0,1,\ldots,K\}\subset\mathcal{X}.
\label{Kdef}
\end{equation}
By filling all the negative energy states, the groundstate $|\Psi\rangle$ is obtained,
\begin{equation}
|\Psi\rangle\eqdef \hat{c}_0^\dagger\hat{c}_1^\dagger\cdots\hat{c}_K^\dagger|0\rangle,
\qquad \langle\Psi|\Psi\rangle=1,
\label{Psidef}
\end{equation}
which is the lowest energy $\sum_{k\in\mathcal{K}}\bigl(\mathcal{E}(k)-\mu\bigr)$ state.
The two point correlation function in the ground state $|\Psi\rangle$ is defined by
\begin{equation}
C_{x,y}\eqdef\langle\Psi|c_x^\dagger c_y|\Psi\rangle,\quad x,y\in\mathcal{X},
\end{equation}
which is the transition probability amplitude between two states $c_y|\Psi\rangle$ and $c_x|\Psi\rangle$.
Expressing $c_x^\dagger$ and $c_y$ by $\{\hat{c}_j^\dagger\}$ and $\{\hat{c}_k\}$ \eqref{cxexp} 
leads to the following
\begin{theo}
\label{theo:gr1}
\begin{equation}
C_{x,y}=\sum_{k\in\mathcal{K}}\hat{\phi}_k(x)\hat{\phi}_k(y)=C_{y,x},\quad x,y\in\mathcal{X},
\label{Csym}
\end{equation}
which means that the eigenvalues are non-negative.
It is obvious that $C_{x,y}$, $x,y\in\mathcal{X}$ has $K+1$-fold eigenvalue $1$  with the eigenvectors 
$\{\hat{\phi}_k(x)\}$, $k\in\mathcal{K}$ and $|\mathcal{X}|-|\mathcal{K}|$-fold eigenvalue $0$ with
the eigenvectors $\{\hat{\phi}_j(x)\}$, $j\in\mathcal{X}\backslash\mathcal{K}$.
\end{theo}
This can be seen easily as
\begin{align*}
C_{x,y}=\langle\Psi|c_x^\dagger c_y|\Psi\rangle&=\sum_{j,k\in\mathcal{X}}\hat{\phi}_j(x)\hat{\phi}_k(y)
\langle\Psi|\hat{c}_j^\dagger\hat{c}_k|\Psi\rangle\\
&=\sum_{j,k\in\mathcal{K}}\hat{\phi}_j(x)\hat{\phi}_k(y)(-1)^{j+k}
\langle\Psi\backslash\{\hat{c}_j\}|\Psi\backslash\{\hat{c}_k^\dagger\}\rangle\\
&=\sum_{j,k\in\mathcal{K}}\hat{\phi}_j(x)\hat{\phi}_k(y)(-1)^{j+k}\delta_{j,k}
=\sum_{k\in\mathcal{K}}\hat{\phi}_k(x)\hat{\phi}_k(y),
\end{align*}
as
\begin{equation*}
\langle\Psi|\hat{c}_j^\dagger=
\left\{
\begin{array}{cc}
0  &   \text{if}\ j\notin\mathcal{K} \\
 (-1)^j\langle\Psi\backslash\{\hat{c}_j\}| & \text{if}\ j\in\mathcal{K}
\end{array}
\right.,
\qquad
\hat{c}_k|\Psi\rangle=
\left\{
\begin{array}{cc}
0  &   \text{if}\ k\notin\mathcal{K} \\
 (-1)^k|\Psi\backslash\{\hat{c}_k^\dagger\}\rangle & \text{if}\ k\in\mathcal{K}
\end{array}
\right..
\end{equation*}
Christoffel-Darboux Theorem \cite{askey}{\bf Th.5.2.4} provides the explicit form of the two point
correlation function $C_{x,y}$ \eqref{Csym} in the following
\begin{theo}
\label{theo:CD}
\begin{align}
C_{x,y}&=\frac{\alpha_K d_K^2\phi_0(x)\phi_0(y)}{\alpha_{K+1}}\cdot
\frac{P_{K+1}\bigl(\eta(x)\bigr)P_{K}\bigl(\eta(y)\bigr)
-P_{K+1}\bigl(\eta(y)\bigr)P_{K}\bigl(\eta(x)\bigr)}{\eta(x)-\eta(y)},
\label{corre2}\\[2pt]
C_{x,x}&=\frac{\alpha_K d_K^2\phi_0(x)\phi_0(y)}{\alpha_{K+1}}\cdot
\Bigl(P'_{K+1}\bigl(\eta(x)\bigr)P_K\bigl(\eta(x)\bigr)-P_{K+1}\bigl(\eta(x)\bigr)P'_K\bigl(\eta(x)\bigr)\Bigr),
\label{corre3}
\end{align}
in which $\alpha_n$ is the coefficient of the highest degree term $\eta(x)^n$ of the polynomial
$P_n\bigl(\eta(x)\bigr)$ and $P'_K\bigl(\eta(x)\bigr)=dP_K\big(\eta(x)\bigr)/d\eta(x)$.
\end{theo}

It is a good challenge to determine the eigenvalues of various submatices of $C_{x,y}$, say
\begin{equation*}
CL_{x,y}=\langle\Psi|c_x^\dagger c_y|\Psi\rangle,\quad x,y\in\mathcal{L}
\eqdef\{0,1,\ldots,L\}\subset\mathcal{X}.
\end{equation*}
They are necessary for the calculation of certain entropy. \cite{ortho,Kra,finkel, latorre}.

\subsection{Fermion Birth and Death operator $\mathcal{L}_{BD}^f$}
\label{sec:FBDeq}

The process can be said quantum fermion walks.
Following the recipe of \eqref{LBDdef2} and \eqref{Hfdef3}, the Birth and Death operator corresponding to the
inhomogeneous fermions is
\begin{align} 
  \mathcal{L}_{BD}^f&\eqdef\sum_{x,y\in\mathcal{X}}( \mathcal{L}_{BD})_{x,y}c_x^\dagger c_y
  \label{fBDdef1}\\
  &=\sum_{x\in\mathcal{X}}\Bigl(B(x-1)c_x^\dagger c_{x-1}+D(x+1)c_x^\dagger c_{x+1}
      -\bigl(B(x)+D(x)\bigr)c_x^\dagger c_x\Bigr),
   \label{fBDdef2}   
\end{align}
and the fermion Birth and Death equation, or quantum fermion walk, reads
\begin{equation}
\frac{\partial|\mathcal{P}(t)\rangle}{\partial t}= \mathcal{L}_{BD}^f|\mathcal{P}(t)\rangle,
\label{fBDeq}
\end{equation}
in which $|\mathcal{P}(t)\rangle$ is a state vector.
It is no longer an equation for a probability distribution. It is a kind of continuous time quantum random walk.
Corresponding to the eigenvectors of $\mathcal{L}_{BD}$, $\{\hat{\phi}_0(x)\hat{\phi}_n(x)\}$ \eqref{LBDeig},
let us introduce
\begin{equation}
\hat{\hat{c}}_n^\dagger\eqdef\sum_{x\in\mathcal{X}}\hat{\phi}_0(x)\hat{\phi}_n(x)c_x^\dagger,
\quad n\in\mathcal{X}.
\label{dhatc}
\end{equation}
Simple calculations lead to the following
\begin{theo}
\label{theo:fBD}
The BD counterparts of {\bf Theorem \ref{thep:main}} read
\begin{equation}
[ \mathcal{L}_{BD}^f,\hat{\hat{c}}_n^\dagger]=-\mathcal{E}(n)\hat{\hat{c}}_n^\dagger,
\quad
[ \mathcal{L}_{BD}^f,\prod_{j\in\mathcal{J}}\hat{\hat{c}}_j^\dagger]=-\sum_{j\in\mathcal{J}}\mathcal{E}(j)\cdot\prod_{j\in\mathcal{J}}\hat{\hat{c}}_j^\dagger,
\label{fBDcomm}
\end{equation}
in which  $\mathcal{J}$ is a subset of $\mathcal{X}$ consisting of mutually distinct elements.
The general solution of the fermion BD equation \eqref{fBDeq} is a linear combination of
\begin{equation}
|\mathcal{P}_{\mathcal J}(t)\rangle
=e^{-\sum_{j\in\mathcal{J}}\mathcal{E}(j)t}\prod_{j\in\mathcal{J}}\hat{\hat{c}}_j^\dagger|0\rangle.
\label{strongdecay}
\end{equation}
This means multi-fermion quantum random walks decay rapidly.
Single fermion BD process is the exception,
\begin{equation}
|\mathcal{P}_S(t)\rangle=\sum_{n\in\mathcal{X}}\beta_n e^{-\mathcal{E}(n)t}\hat{\hat{c}}_n^\dagger|0\rangle
\stackrel{t\to+\infty}{\longrightarrow}\beta_0\hat{\hat{c}}_0^\dagger|0\rangle
=\beta_0\sum_{x\in\mathcal{X}}\hat{\phi}_0^2(x)c_x^\dagger|0\rangle,
\end{equation}
approaching to the stationary distribution.
A single fermion BD process starting from $y\in\mathcal{X}$ is 
\begin{equation}
|\mathcal{P}_y(t)\rangle=\hat{\phi}_0^{-1}(y)\sum_{n\in\mathcal{X}}\hat{\phi}_n(y)e^{-\mathcal{E}(n)t}\hat{\hat{c}}_n^\dagger|0\rangle=\sum_{x,n\in\mathcal{X}}\hat{\phi}_0(x)\hat{\phi}_0(y)^{-1}\hat{\phi}_n(x)\hat{\phi}_n(y)e^{-\mathcal{E}(n)t}c_x^\dagger|0\rangle,
\label{bdsol2}
\end{equation}
which looks similar to the classical case \eqref{bdsoly1}.
\end{theo}

\section{Exactly solvable XX spin models }
\label{sec:solvspin}

It is well known that XX spin chains and nearest neighbour interacting fermion systems are connected
by the Jordan-Wigner transformation.
Let $\sigma_x^{i}$, $i=1,2,3$ be the Pauli spin matrices on the lattice point $x\in\mathcal{X}$,
\begin{equation*}
\sigma^1=
\left(
\begin{array}{cc}
0  &  1 \\
1  &   0   
\end{array}
\right), \ 
\sigma^2=
\left(
\begin{array}{cc}
0  &  -i \\
i  &   0   
\end{array}
\right), \
\sigma^3=
\left(
\begin{array}{cc}
1  &  0 \\
0  &  -1   
\end{array}
\right), \
\sigma^+=
\left(
\begin{array}{cc}
0  &  1 \\
0  &  0   
\end{array}
\right), \
\sigma^-=
\left(
\begin{array}{cc}
0  &  0 \\
1  &  0   
\end{array}
\right).
\end{equation*}
Defining $c_x^\dagger$ and $c_x$, $x\in\mathcal{X}$, in terms of the lattice Paui matrices 
 by Jordan-Wigner transformation
\begin{equation}
c_x^\dagger\eqdef \sigma_0^3\cdots \sigma_{x-1}^3\sigma_x^{+},\quad
c_x= \sigma_0^3\cdots \sigma_{x-1}^3\sigma_x^{-}, \quad x\in\mathcal{X};
\ \Rightarrow c_x^\dagger c_x=\frac12(1+\sigma_x^3),
\label{JWrel}
\end{equation}
it is easy to verify that they satisfy the fermion anti-commutation relations \eqref{comrel}.
The corresponding vacuum of the spin system is 
\begin{equation}
\sigma_x^-|0\rangle_s=0,\quad \forall x\in\mathcal{X},\quad 
|0\rangle_s=\frac1{\sqrt{|\mathcal{X}|}}\prod_{x\in\mathcal{X}}\left(
\begin{array}{c}
0   \\
1  
\end{array}
\right)_x,\quad {}_s\langle0|0\rangle_s=1.
\end{equation}
In fact, for the actual calculations, the normalisation condition ${}_s\langle0|0\rangle_s=1$
and commutation relations of Pauli matrices are needed, as in the fermion cases. 
Therefore, finite $|\mathcal{X}|$ condition is not necessary.

\subsection{Exactly solvable spin Hamiltonian}
\label{sec:solvHamspin}
By rewriting the fermion Hamiltonian $\mathcal{H}_f$ \eqref{Hfdef1} 
in terms of the Jordan-Wigner transformation \eqref{JWrel}, the XX spin Hamiltonian is obtained
\begin{align}
\mathcal{H}_s&=\frac12\sum_{x\in\mathcal{X}}
\left\{\sqrt{B(x)D(x+1)}\bigl(\sigma_x^1 \sigma_{x+1}^1+\sigma_{x}^2 \sigma_{x+1}^2\bigr)
+\bigl(B(x)+D(x)-\mu\bigr)\bigl(1+\sigma_x^3\bigr)\right\},
\label{Hsdef1}\\
&=\sum_{x\in\mathcal{X}}
\left\{\sqrt{B(x)D(x+1)}\bigl(\sigma_x^+ \sigma_{x+1}^-+\sigma_{x}^- \sigma_{x+1}^+\bigr)
+\tfrac12\bigl(B(x)+D(x)-\mu\bigr)\bigl(1+\sigma_x^3\bigr)\right\},
\label{Hsdef12}\\
&=\sum_{x\in\mathcal{X}}
\left\{\sqrt{B(x)D(x+1)}\,\sigma_x^+ \sigma_{x+1}^-+\sqrt{B(x-1)D(x)}\,\sigma_x^+ \sigma_{x-1}^-\right.\n
&\qquad \qquad +\bigl(B(x)+D(x)-\mu\bigr)\frac12\bigl(1+\sigma_x^3\bigr)\Bigr\},
\label{Hsdef13}
\end{align}
as $c_x^\dagger c_{x+1}=-\sigma_x^+\sigma_{x+1}^-$ ans $c_{x+1}^\dagger c_x=-\sigma_x^-\sigma_{x+1}^+$.
In this section, $\mu$ is a constant magnetic field.
Corresponding to the fermion number conservation \eqref{fcons} and \eqref{JWrel}, 
total number of up spin states $\mathcal{S}$
is conserved
\begin{equation}
[\mathcal{H}_s,\mathcal{S}]=0,\qquad \mathcal{S}\eqdef \frac12\sum_{x\in\mathcal{X}}(1+\sigma_x^3),
\qquad \mathcal{H}_s|0\rangle_s=0.
\label{scons}
\end{equation}
The Schr\"odinger equation reads
\begin{equation}
i\frac{\partial|\psi_s(t)\rangle}{\partial t}=\mathcal{H}_s|\psi_s(t)\rangle,
\label{Scheqs}
\end{equation}
in which $|\psi_s(t)\rangle$ is a state vector.
The eigenvectors of the XX spin Hamiltonian $\mathcal{H}_s$ \eqref{Hsdef1} are grouped according to the
number of up spin states.
By defining 
\begin{equation}
\hat{s}_n\eqdef \sum_{x\in\mathcal{X}}\hat{\phi}_n(x)c_x
=\sum_{x\in\mathcal{X}}\hat{\phi}_n(x)\sigma_0^3\cdots \sigma_{x-1}^3\sigma_x^{-},
\quad \hat{s}_n^\dagger=\sum_{x\in\mathcal{X}}\hat{\phi}_n(x)\sigma_0^3\cdots \sigma_{x-1}^3\sigma_x^{+},
\label{sndef}
\end{equation}
one arrives at the following

\begin{theo}
\label{normindep}
\label{theo:smain}
Fifteen XX spin Hamiltonians \eqref{Hsdef1} associated with the discrete orthogonal polynomials of Askey scheme are
exactly solvable
\begin{align}
&\hspace{5cm}\mathcal{H}_{s}=\sum_{n\in\mathcal{X}}\bigl(\mathcal{E}(n)-\mu\bigr)\hat{s}_n^\dagger \hat{s}_n
\label{Hsdef2}\\
&
\quad [\mathcal{H}_{s},\hat{s}_n^\dagger]=\bigl(\mathcal{E}(n)-\mu\bigr)\hat{s}_n^\dagger,
\qquad [\mathcal{H}_{s},\hat{s}_n]=-\bigl(\mathcal{E}(n)-\mu\bigr)\hat{s}_n,
\label{Hscom1}\\[2pt]
&
\quad [\mathcal{H}_{s},\prod_{j\in\mathcal{J}}\hat{s}_j^\dagger]
=\sum_{j\in\mathcal{J}}\bigl(\mathcal{E}(j)-\mu\bigr)\cdot\prod_{j\in\mathcal{J}}\hat{s}_j^\dagger,
\qquad [\mathcal{H}_{s},\prod_{j\in\mathcal{J}}\hat{s}_j]
=-\sum_{j\in\mathcal{J}}\bigl(\mathcal{E}(j)-\mu\bigr)\cdot\prod_{j\in\mathcal{J}}\hat{s}_j
\label{HscomJ},
\end{align}
in which, as before, $\mathcal{J}$ is a subset of $\mathcal{X}$ consisting of mutually distinct elements.
As before \eqref{eigst},
the  general eigenstate {\rm(}eigenvector\/{\rm)} of $\mathcal{H}_s$ \eqref{Hsdef1} with $|\mathcal{J}|$ up spins,  has the form 
\begin{align}
\prod_{j\in\mathcal{J}}\hat{s}_j^\dagger|0\rangle_s,\quad 
\mathcal{H}_s\prod_{j\in\mathcal{J}}\hat{s}_j^\dagger|0\rangle_s
&=E\prod_{j\in\mathcal{J}}\hat{s}_j^\dagger|0\rangle_s
\quad E=\sum_{j\in\mathcal{J}}\bigl(\mathcal{E}(j)-\mu\bigr),
\label{seigst}\\
\mathcal{S}\prod_{j\in\mathcal{J}}\hat{s}_j^\dagger|0\rangle_s
&=|\mathcal{J}|\prod_{j\in\mathcal{J}}\hat{s}_j^\dagger|0\rangle_s.
\label{Seigst}
\end{align}
The general $|\mathcal{J}|$ up spin   solution of the Schr\"odinger equation \eqref{Scheqs} is a linear 
combination of 
\begin{equation}
|\psi_{s\mathcal{J}}(t)\rangle
= e^{-i\sum_{j\in\mathcal{J}}(\mathcal{E}(n)-\mu)t}\prod_{j\in\mathcal{J}}\hat{s}_j^\dagger|0\rangle_s,
\label{sJgex}
\end{equation}
in which $\mathcal{J}$ is a subset of $\mathcal{X}$ consisting of distinct elements.
The general single up spin excitation solution of the spin Schr\"odinger equation \eqref{Scheqs} reads
\begin{align}
|\psi_{s1}(t)\rangle&=\sum_{n\in\mathcal{X}}\beta_n e^{-i(\mathcal{E}(n)-\mu)t}\hat{s}_n^\dagger|0\rangle_s
=\sum_{x,n\in\mathcal{X}}\beta_n\hat{\phi}_n(x)e^{-i(\mathcal{E}(n)-\mu)t}
\sigma_0^3\cdots \sigma_{x-1}^3\sigma_x^{+}|0\rangle_s\n
&=\sum_{x,n\in\mathcal{X}}\beta_n\hat{\phi}_n(x)e^{-i(\mathcal{E}(n)-\mu)t}(-1)^x\sigma_x^{+}|0\rangle_s,
\qquad \beta_n\in\mathbb{C}.
\label{ssingex}
\end{align}
\end{theo}
\begin{rema}
\label{altspin}
If the parity oscillating factor $(-1)^x$ in the single up spin excitation solution \eqref{ssingex} is cumbersome,
one could have started from the alternative fermion Hamiltonian $\mathcal{H}_{f'}$ \eqref{Hfpdef} leading to 
the alternative XX spin Hamiltonian $\mathcal{H}_{s'}$
\begin{align}
\hspace{-3mm}\mathcal{H}_{s'}&=\frac12\sum_{x\in\mathcal{X}}
\left\{-\sqrt{B(x)D(x+1)}\bigl(\sigma_x^1 \sigma_{x+1}^1+\sigma_{x}^2 \sigma_{x+1}^2\bigr)
+\bigl(B(x)+D(x)-\mu\bigr)\bigl(1+\sigma_x^3\bigr)\right\},
\label{Hspdef1}\\
&=\sum_{x\in\mathcal{X}}
\left\{-\sqrt{B(x)D(x+1)}\bigl(\sigma_x^+ \sigma_{x+1}^-+\sigma_{x}^- \sigma_{x+1}^+\bigr)
+\tfrac12\bigl(B(x)+D(x)-\mu\bigr)\bigl(1+\sigma_x^3\bigr)\right\},
\label{Hspdef12}\\
&=\sum_{x\in\mathcal{X}}
\left\{-\sqrt{B(x)D(x+1)}\,\sigma_x^+ \sigma_{x+1}^--\sqrt{B(x-1)D(x)}\,\sigma_x^+ \sigma_{x-1}^-\right.\n
&\qquad \qquad +\bigl(B(x)+D(x)-\mu\bigr)\frac12\bigl(1+\sigma_x^3\bigr)\Bigr\},
\label{Hspdef13}
\end{align}
\end{rema}

\bigskip
It is almost straightforward to write down the spin version of the contents in \S\ref{sec:ground} and \S\ref{sec:FBDeq}.
Probably they are easier to handle when the alternative spin Hamiltonian 
$\mathcal{H}_{s'}$ \eqref{Hspdef1}--\eqref{Hspdef13} is adopted.
As the expressions \eqref{scons} to \eqref{ssingex} show that they are very similar to the fermion counterparts
and I do not present them here.

\section*{Acknowledgements}
R.\,S. thanks Junji Suzuki for useful discussions.

\appendix
\section*{Appendix:Data}
\label{append}
\setcounter{equation}{0}
\renewcommand{\theequation}{A.\arabic{equation}}
\renewcommand{\thesubsection}{A.\arabic{subsection}}

In Appendix, the necessary data for the evaluation of various quantities  of 
the fifteen discrete orthogonal polynomials are provided.
They are $B(x)$ and $D(x)$ for the definition of the Hamiltonian 
$\mathcal{H}^A$ and for specifying the parameter ranges, 
the sinusoidal coordinate $\eta(x)$, the energy eigenvalue $\mathcal{E}(n)$, 
the polynomial $\check{P}_n(x)=P_n\bigl(\eta(x)\bigr)$,
the orthogonality measure
$\phi_0^2(x)$, the normalisation constant $d_n^2$ and the highest degree coefficient $\alpha_n$.
Ten polynomials defined on a finite lattice are followed by five polynomials on a semi-infinite lattice.
For the full details of these exactly solvable discrete quantum systems,
a paper by Odake and myself \cite{os12}
should be consulted.

\subsection{Finite polynomials}
\paragraph{Krawtchouk} \quad ($0<p<1$)
\label{sec:Kr}
\begin{align*}
&B(x)=p(N-x),\quad D(x)=(1-p)x,\quad 0<p<1,\quad \eta(x)=x,\quad \mathcal{E}(n)=n,\\
&     {P}_n(x)={}_2F_1\Bigl(
  \genfrac{}{}{0pt}{}{-n,\,-x}{-N}\Bigm|p^{-1}\Bigr),\quad  \phi_0(x)^2=
  \frac{N!}{x!\,(N-x)!}\Bigl(\frac{p}{1-p}\Bigr)^x,\\
&  d_n^2
  =\frac{N!}{n!\,(N-n)!}\Bigl(\frac{p}{1-p}\Bigr)^n\times(1-p)^N,\quad
\alpha_n=  (-1)^n\frac{p^{-n}(N-n)!}{N!}.
\end{align*}
\paragraph{Hahn} \quad ($a,b>0$)
\begin{align*}
  &B(x)=(x+a)(N-x),\quad D(x)= x(b+N-x),\quad \eta(x)=x,\quad
\mathcal{E}(n)=n(n+a+b-1),\\[2pt]
  &{P}_n(x)
  ={}_3F_2\Bigl(\genfrac{}{}{0pt}{}{-n,\,n+a+b-1,\,-x}{a,\,-N}\Bigm|1\Bigr),\quad
  \phi_0(x)^2
 =\frac{N!}{x!\,(N-x)!}\,\frac{(a)_x\,(b)_{N-x}}{(b)_N},
 \\[2pt]
&d_n^2
  =\frac{N!}{n!\,(N-n)!}\,
  \frac{(a)_n\,(2n+a+b-1)(a+b)_N}{(b)_n\,(n+a+b-1)_{N+1}}
  \times\frac{(b)_N}{(a+b)_N}, \\[4pt]
& \alpha_n=(-1)^n\frac{(n+a+b-1)_n(N-n)!}{(a)_nN!}.
\end{align*}
\paragraph{Dual Hahn} ($a,b>0$)
\begin{align*}
&B(x)=\frac{(x+a)(x+a+b-1)(N-x)}
  {(2x-1+a+b)(2x+a+b)},
\quad
D(x)=\frac{x(x+b-1)(x+a+b+N-1)}
  {(2x-2+a+b)(2x-1+a+b)},\\[2pt]
  &\eta(x)=x(x+a+b-1),\quad \mathcal{E}(n)=n,\quad
 {P}_n\bigl(\eta(x)\bigr)\!
  ={}_3F_2\Bigl(\genfrac{}{}{0pt}{}{-n,\,x+a+b-1,\,-x}{a,\,-N}\Bigm|1\Bigr),\\[2pt] 
&    \phi_0(x)^2
 \! =\!\frac{N!}{x!\,(N-x)!}
  \frac{(a)_x\,(2x+a+b-1)(a+b)_N}{(b)_x\,(x+a+b-1)_{N+1}},\\[2pt]
  &d_n^2
  =\frac{N!}{n!\,(N-n)!}\,\frac{(a)_n\,(b)_{N-n}}{(b)_N}
  \times\frac{(b)_{N}}{(a+b)_N}, \quad \alpha_n=(-1)^n\frac{(N-n)!}{(a)_nN!}.
\end{align*}
\paragraph{Racah} \quad  ($d>0,\ a>N+d,\ 0<b<1+d$)
\begin{align*}
&B(x)
  =-\frac{(x+a)(x+b)(x-N)(x+d)}{(2x+d)(2x+d+1)},\quad
D(x)
  =-\frac{(x+d-a)(x+d-b)(x+d+N)x}{(2x+d-1)(2x+d)},\\[2pt]
&\eta(x)=x(x+d),\quad \mathcal{E}(n)=n(n+\tilde{d}),\quad \tilde{d}\eqdef a+b-N-d-1,\\[2pt]
  &P_n\bigl(\eta(x)\bigr)
  ={}_4F_3\Bigl(
  \genfrac{}{}{0pt}{}{-n,\,n+\tilde{d},\,-x,\,x+d}
  {a,\,b,\,-N}\Bigm|1\Bigr),\\[2pt]
&  \phi_0(x)^2=\frac{(a,b,-N,d)_x}{(1+d-a,1+d-b,1+d+N,1)_x}\,
  \frac{2x+d}{d},\\[2pt]
&d_n^2=\frac{(a,b,-N,\tilde{d})_n}
  {(1+\tilde{d}-a,1+\tilde{d}-b,1+\tilde{d}+N,1)_n}\,
  \frac{2n+\tilde{d}}{\tilde{d}} \\[2pt]
  &\qquad\qquad\quad\times
  \frac{(-1)^N(1+d-a,1+d-b,1+d+N)_N}{(\tilde{d}+1)_N(d+1)_{2N}},\quad
  \alpha_n=(-1)^n\frac{(n+\tilde{d})_n(N-n)!}{(a)_n(b)_nN!}.
\end{align*}
\paragraph{quantum $q$-Krawtchouk}\ ($p>q^{-N}$)
\begin{align*}
&B(x)=p^{-1}q^x(q^{x-N}-1),\
  D(x)=(1-q^x)(1-p^{-1}q^{x-N-1}),\  \eta(x)=q^{-x}-1,\
\mathcal{E}(n)=1-q^n,\\[2pt]
 &P_n\bigl(\eta(x)\bigr)
  ={}_2\phi_1\Bigl(
  \genfrac{}{}{0pt}{}{q^{-n},\,q^{-x}}{q^{-N}}\Bigm|q\,;pq^{n+1}\Bigr),\quad
  \phi_0(x)^2
  =\frac{(q\,;q)_N}{(q\,;q)_x(q\,;q)_{N-x}}\,
  \frac{p^{-x}q^{x(x-1-N)}}{(p^{-1}q^{-N}\,;q)_x},\\[2pt]
&d_n^2
  =\frac{(q\,;q)_N}{(q\,;q)_n(q\,;q)_{N-n}}\,
  \frac{p^{-n}q^{-Nn}}{(p^{-1}q^{-n}\,;q)_n}\,
  \times(p^{-1}q^{-N}\,;q)_N,\quad \alpha_n=(-1)^n\frac{p^n q^{n(N+1)}(q;q)_{N-n}}{(q;q)_N}.
\end{align*}
                                                %
 \paragraph{$q$-Krawtchouk} ($p>0$)
\begin{align*}
&B(x)=q^{x-N}-1,\quad
  D(x)=p(1-q^x),\quad
\mathcal{E}(n)=(q^{-n}-1)(1+pq^n),\quad
  \eta(x)=q^{-x}-1,\\[2pt]
&P_n\bigl(\eta(x)\bigr)
  ={}_3\phi_2\Bigl(
  \genfrac{}{}{0pt}{}{q^{-n},\,q^{-x},\,-pq^n}{q^{-N},\,0}\Bigm|q\,;q\Bigr),\quad
  \phi_0(x)^2=\frac{(q\,;q)_N}{(q\,;q)_x(q\,;q)_{N-x}}\,
  p^{-x}q^{\frac12x(x-1)-xN},\\[2pt]
 &d_n^2
  =\frac{(q\,;q)_N}{(q;q)_n(q;q)_{N-n}}\,
  \frac{(-p\,;q)_n}{(-pq^{N+1}\,;q)_n\,p^nq^{\frac12n(n+1)}}\,
  \frac{1+pq^{2n}}{1+p}
  \times\frac{p^{N}q^{\frac12N(N+1)}}{(-pq\,;q)_N},\\[2pt]
&\alpha_n=(-1)^n q^{n(N-n+1)}\frac{(-pq^n;q)_n(q;q)_{N-n}}{(q;q)_N}.
\end{align*}

                                                   %
\paragraph{affine $q$-Krawtchouk} ($0<p<q^{-1}$)
\begin{align*}
& B(x)=(q^{x-N}-1)(1-pq^{x+1}),\
  D(x)=pq^{x-N}(1-q^x),\
\mathcal{E}(n)=q^{-n}-1,\   \eta(x)=q^{-x}-1,\\[2pt]
& P_n\bigl(\eta(x)\bigr)
 ={}_3\phi_2\Bigl(
  \genfrac{}{}{0pt}{}{q^{-n},\,q^{-x},\,0}{pq,\,q^{-N}}\Bigm|q\,;q\Bigr),\quad
  \phi_0(x)^2=\frac{(q\,;q)_N}{(q\,;q)_x(q\,;q)_{N-x}} \frac{(pq\,;q)_x}{(pq)^x},\\[2pt]
&d_n^2
  =\frac{(q\,;q)_N}{(q\,;q)_n(q\,;q)_{N-n}}\,
  \frac{(pq\,;q)_n}{(pq)^n}\times(pq)^N,\quad \alpha_n=(-1)^n\frac{q^{n(N-n+1)}(q;q)_{N-n}}{(pq;q)_n(q;q)_N}.
\end{align*}
 \paragraph{$q$-Hahn} ($0<a,b<1$)
\begin{align*}
&B(x) =(1-aq^x)(q^{x-N}-1),\quad
  D(x)= aq^{-1}(1-q^x)(q^{x-N}-b),\\
  &\eta(x)=q^{-x}-1,\quad \mathcal{E}(n)
  =(q^{-n}-1)(1-abq^{n-1}),\\[2pt]
& P_n\bigl(\eta(x)\bigr)
  ={}_3\phi_2\Bigl(
  \genfrac{}{}{0pt}{}{q^{-n},\,abq^{n-1},\,q^{-x}}
  {a,\,q^{-N}}\Bigm|q\,;q\Bigr),\quad \phi_0(x)^2
  =\frac{(q\,;q)_N}{(q\,;q)_x\,(q\,;q)_{N-x}}\,
  \frac{(a;q)_x\,(b\,;q)_{N-x}}{(b\,;q)_N\,a^x},\\[2pt]
&d_n^2
  =\frac{(q\,;q)_N}{(q\,;q)_n\,(q\,;q)_{N-n}}\,
  \frac{(a,abq^{-1};q)_n}{(abq^N,b\,;q)_n\,a^n}\,
  \frac{1-abq^{2n-1}}{1-abq^{-1}}
  \times\frac{(b\,;q)_N\,a^N}{(ab\,;q)_N},\\[2pt]
&\alpha_n=(-1)^n q^{n(N-n+1)}\frac{(abq^n;q)_n(q;q)_{N-n}}{(a;q)_n(q;q)_N}.
\end{align*}
\paragraph{dual $q$-Hahn} ($0<a,b<1$)
\begin{align*}
&B(x)=
  \frac{(q^{x-N}-1)(1-aq^x)(1-abq^{x-1})}
  {(1-abq^{2x-1})(1-abq^{2x})},\\[2pt]
&D(x)=aq^{x-N-1}
  \frac{(1-q^x)(1-abq^{x+N-1})(1-bq^{x-1})}
  {(1-abq^{2x-2})(1-abq^{2x-1})},\\[2pt]
&\mathcal{E}(n)=q^{-n}-1,\quad
  \eta(x)=(q^{-x}-1)(1-abq^{x-1}),
\ \ P_n\bigl(\eta(x)\bigr)
  ={}_3\phi_2\Bigl(
  \genfrac{}{}{0pt}{}{q^{-n},\,abq^{x-1},\,q^{-x}}
  {a,\,q^{-N}}\Bigm|q\,;q\Bigr),\\[2pt]
 &\phi_0(x)^2
  =\frac{(q\,;q)_N}{(q\,;q)_x\,(q\,;q)_{N-x}}\,
  \frac{(a,abq^{-1}\,;q)_x}{(abq^N,b\,;q)_x\,a^x}\,
  \frac{1-abq^{2x-1}}{1-abq^{-1}},\\[2pt]
 &d_n^2
  =\frac{(q\,;q)_N}{(q\,;q)_n\,(q\,;q)_{N-n}}\,
  \frac{(a\,;q)_n(b\,;q)_{N-n}}{(b;q)_N\,a^n}
  \times\frac{(b\,;q)_N\,a^N}{(ab;q)_N},\quad 
  \alpha_n=(-1)^n \frac{q^{n(N-n+1)}(q;q)_{N-n}}{(a;q)_n(q;q)_N}.
\end{align*}

\paragraph{$q$-Racah}\ ($0<d<1,\ 0<a<q^Nd,\ qd<b<1$)
\begin{align*}
&B(x) =-\frac{(1-aq^x)(1-bq^x)(1-q^{x-N})(1-dq^x)}
  {(1-dq^{2x})(1-dq^{2x+1})},\\[2pt]
&  D(x)= -\tilde{d}\,
  \frac{(1-a^{-1}dq^x)(1-b^{-1}dq^x)(1-dq^{N+x})(1-q^x)}
  {(1-dq^{2x-1})(1-dq^{2x})},\\[2pt]
 &\mathcal{E}(n)=(q^{-n}-1)(1-\tilde{d}q^n),\quad
  \eta(x)=(q^{-x}-1)(1-dq^x),\quad \tilde{d}\eqdef abd^{-1}q^{-N-1},\\[2pt]
  &P_n\bigl(\eta(x)\bigr)
  ={}_4\phi_3\Bigl(
  \genfrac{}{}{0pt}{}{q^{-n},\,\tilde{d}q^n,\,q^{-x},\,dq^x}
  {a,\,b,\,q^{-N}}\Bigm|q\,;q\Bigr),\\[2pt]
   &\phi_0(x)^2=\frac{(a,b,q^{-N},d\,;q)_x}
  {(a^{-1}dq,b^{-1}dq,dq^{N+1},q\,;q)_x\,\tilde{d}^x}\,
  \frac{1-dq^{2x}}{1-d}\,,\\[2pt]
  &d_n(\bm{\lambda})^2
  =\frac{(a,b,q^{-N},\tilde{d}\,;q)_n}
  {(a^{-1}\tilde{d}q,b^{-1}\tilde{d}q,\tilde{d}q^{N+1},q\,;q)_n\,d^n}\,
  \frac{1-\tilde{d}q^{2n}}{1-\tilde{d}}\\
  &\qquad\qquad\quad\times
  \frac{(-1)^N(a^{-1}dq,b^{-1}dq,dq^{N+1}\,;q)_N\,\tilde{d}^Nq^{\frac12N(N+1)}}
  {(\tilde{d}q\,;q)_N(dq\,;q)_{2N}},\\
  &\alpha_n=(-1)^n \frac{q^{n(N-n+1)}(\tilde{d}q^n;q)_n(q;q)_{N-n}}{(a;q)_n(b;q)_n(q;q)_N}.
\end{align*}

\subsection{Infinite polynomials}
\paragraph{Meixner} ($\beta>0$, \ $0<c<1$)
\begin{align*}
 & B(x)=\frac{c}{1-c}(x+b),\quad
  D(x)=\frac{x}{1-c},\quad \eta(x)=x,\quad \mathcal{E}(n)=n,\\
 &P_n(x)
  ={}_2F_1\Bigl(
  \genfrac{}{}{0pt}{}{-n,\,-x}{\beta}\Bigm|1-c^{-1}\Bigr),  \quad
\phi_0(x)^2=\frac{(\beta)_x\,c^x}{x!},\quad
  d_n^2
  =\frac{(\beta)_n\,c^n}{n!}\times(1-c)^{\beta},\\
& \alpha_n=\frac{(1-c^{-1})^n}{(\beta)_n}.
\end{align*}
\paragraph{Charlier}  ($a>0$)
\begin{align*}
& B(x)=a,\quad D(x)=x,\quad \eta(x)=x,\quad \mathcal{E}(n)=n,\quad \alpha_n=(-a)^{-n},\\
 &P_n(x)
  ={}_2F_0\Bigl(
  \genfrac{}{}{0pt}{}{-n,\,-x}{-}\Bigm|-a^{-1}\Bigr),\quad
\phi_0(x)^2=\frac{a^x}{x!},\quad
  d_n^2
  =\frac{a^{n}}{n!}\times e^{-a}.
\end{align*}
\paragraph{Little $q$-Jacobi}  ($0<a<q^{-1}$, $b<q^{-1}$)
\begin{align*}
 &B(x)=a(q^{-x}-bq),\quad
  D(x)=q^{-x}-1,\quad  \eta(x)=1-q^x, \quad 
\mathcal{E}(n)=(q^{-n}-1)(1-abq^{n+1}),\\[2pt]
  &P_n\bigl(\eta(x)\bigr)
  =(-a)^{-n}q^{-\frac12n(n+1)}\frac{(aq\,;q)_n}{(bq\,;q)_n}\,
  {}_2\phi_1\Bigl(
  \genfrac{}{}{0pt}{}{q^{-n},\,abq^{n+1}}{aq}\Bigm|q\,;q^{x+1}\Bigr)\\[2pt]
  &\phi_0(x)^2=\frac{(bq\,;q)_x}{(q\,;q)_x}(aq)^x,\quad
  d_n^2
  =\frac{(bq,abq\,;q)_n\,a^nq^{n^2}}{(q,aq\,;q)_n}\,
  \frac{1-abq^{2n+1}}{1-abq}
  \times\frac{(aq\,;q)_{\infty}}{(abq^2\,;q)_{\infty}},\\
 & \alpha_n=(-a)^{-n}q^{-n^2}\frac{(abq^{n+1};q)_n}{(bq;q)_n}.
\end{align*}
\paragraph{Little $q$-Laguerre}  ($0<a<q^{-1}$)
\begin{align*} 
  &B(x)=aq^{-x},\quad
  D(x)=q^{-x}-1,\quad  \eta(x)=1-q^x,\quad 
\mathcal{E}(n)=q^{-n}-1, \\
  &P_n\bigl(\eta(x)\bigr)
  ={}_2\phi_0\Bigl(
  \genfrac{}{}{0pt}{}{q^{-n},\,q^{-x}}{-}\Bigm|q\,;a^{-1}q^x\Bigr),
\quad
\phi_0(x)^2=\frac{(aq)^x}{(q\,;q)_x},\\
&  d_n^2
  =\frac{a^nq^{n^2}}{(q,aq\,;q)_n}\times(aq\,;q)_{\infty},\quad
  \alpha_n=(-a)^{-n}q^{-n^2}.
  \end{align*}
\paragraph{Al-Salam-Carlitz II}  ($0<a<q^{-1}$)
\begin{align*} 
 &B(x)=aq^{2x+1},\quad
  D(x)=(1-q^x)(1-aq^x), \quad  \eta(x)=q^{-x}-1, \quad
\mathcal{E}(n)=1-q^n,\\
  &P_n\bigl(\eta(x)\bigr)
  ={}_2\phi_0\Bigl(
  \genfrac{}{}{0pt}{}{q^{-n},\,q^{-x}}{-}\Bigm|q\,;a^{-1}q^n\Bigr),\quad
 \phi_0(x)^2=\frac{a^xq^{x^2}}{(q,aq\,;q)_x},\\
&  d_n^2
  =\frac{(aq)^n}{(q\,;q)_n}\times(aq\,;q)_{\infty},\quad \alpha_n=(-a)^{-n}.
  \end{align*}

\end{document}